\documentclass[conference]{IEEEtran}
\IEEEoverridecommandlockouts
\usepackage{cite}
\usepackage{amsmath,amssymb,amsfonts}
\usepackage{algorithmic}
\usepackage{graphicx}
\usepackage{textcomp}
\usepackage{xcolor}

\def\BibTeX{{\rm B\kern-.05em{\sc i\kern-.025em b}\kern-.08em
    T\kern-.1667em\lower.7ex\hbox{E}\kern-.125emX}}

\usepackage{multirow}
\usepackage{float}
\begin{document}

\title{Contracting Tsetlin Machine with\\Absorbing Automata
\thanks{This work is supported in part by the project ``Logic-based Artificial Intelligence Everywhere: Tsetlin Machines in Hardware'' and funded under grant number 312434 of the Research Council of Norway.}
}

\author{\IEEEauthorblockN{Bimal Bhattarai}
\IEEEauthorblockA{\textit{Centre for AI Research} \\
\textit{University of Agder}\\
Grimstad, Norway \\
bimal.bhattarai@uia.no}
\and
\IEEEauthorblockN{Ole-Christoffer Granmo}
\IEEEauthorblockA{\textit{Centre for AI Research} \\
\textit{University of Agder}\\
Grimstad, Norway \\
0000-0002-7287-030X}
\and
\IEEEauthorblockN{Lei Jiao}
\IEEEauthorblockA{\textit{Centre for AI Research} \\
\textit{University of Agder}\\
Grimstad, Norway \\
0000-0002-7115-6489}
\and
\IEEEauthorblockN{Per-Arne Andersen}
\IEEEauthorblockA{\textit{Centre for AI Research} \\
\textit{University of Agder}\\
Grimstad, Norway \\
0000-0002-5490-6436}
\and
\IEEEauthorblockN{Svein Anders Tunheim}
\IEEEauthorblockA{\textit{Centre for AI Research} \\
\textit{University of Agder}\\
Grimstad, Norway \\
0000-0001-7947-8485}
\and
\IEEEauthorblockN{Rishad Shafik}
\IEEEauthorblockA{\textit{Microsystems Group, School of Engineering} \\
\textit{Newcastle University}\\
Newcastle upon Tyne, UK \\
0000-0001-5444-537X}
\and
\IEEEauthorblockN{Alex Yakovlev}
\IEEEauthorblockA{\textit{Microsystems Group, School of Engineering} \\
\textit{Newcastle University}\\
Newcastle upon Tyne, UK \\
0000-0003-0826-9330}
}

\maketitle

\begin{abstract}
In this paper, we introduce a sparse Tsetlin Machine (TM) with absorbing Tsetlin Automata (TA) states. In brief, the TA of each clause literal has both an absorbing Exclude- and an absorbing Include state, making the learning scheme absorbing instead of ergodic. When a TA reaches an absorbing state, it will never leave that state again. If the absorbing state is an Exclude state, both the automaton and the literal can be removed from further consideration. The literal will as a result never participates in that clause. If the absorbing state is an Include state, on the other hand, the literal is stored as a permanent part of the clause while the TA is discarded. A novel sparse data structure supports these updates by means of three action lists: Absorbed Include, Include, and Exclude. By updating these lists, the TM gets smaller and smaller as the literals and their TA withdraw. In this manner, the computation accelerates during learning, leading to faster learning and less energy consumption.
\end{abstract}


\section{Introduction}

The Tsetlin Machine~(TM) is a pattern recognition approach that employs a group of Tsetlin Automata~(TA) to produce logical rules by formulating conjunctive clauses. In brief, the TM unifies summation-based (cf. logistic regression) and rule-based approaches (cf. decision trees), enabling online learning of non-linear patterns akin to neural network learning. However, while deep learning networks are non-transparent making them difficult to interpret~\cite{mengnan}, TMs express patterns logically in sparse disjunctive normal form. As such, they are easily interpretable by humans. Furthermore, the TM's logical representation makes it hardware-friendly, yielding a low energy footprint~\cite{wheeldon2020learning}.

TMs form a versatile framework that can accommodate multiple machine learning paradigms, including classification~\cite{granmo2018tsetlin}, convolution~\cite{granmo2019convtsetlin}, regression~\cite{abeyrathna2020nonlinear}, autoencoding\cite{bhattaray2023embedding}, and reinforcement learning~\cite{DBLP:journals/apin/GorjiG23,RaihanNIPS22}. Its parallel architecture further facilitates constant-time scaling through GPU-based acceleration~\cite{abeyrathna2020massively}. Recent research explores diverse application of TMs, such as sentiment analysis~\cite{rohan2021AAAI}, novelty detection~\cite{bhattarai2021word,bhattarai2021measuring}, fake news detection~\cite{fakenewsLREC}, counterfactual robustness~\cite{yadav2022robustness}, and representation learning~\cite{bhattarai2023interpretable}.

While the TM performs competitively with deep learning state-of-the-art models in several application domains, the computation time required for training remains significant. Research in~\cite{abeyrathna2020massively} introduced a parallel GPU-based architecture, exploiting the inherent parallelism of TM learning. However, since TA learning is ergodic by nature, a TA never stops learning. The current TM learning schemes thus appear sub-optimal. Indeed, it appears advantageous to find a way to stop TA learning as soon as the TA meets its goal, cutting down on computation time and energy consumption. 

Indeed, reducing energy consumption during training is desirable for all machine learning solutions, whether they are deployed in high-end data centers or as low-end battery-operated systems. Typically, the energy consumption of a TM-based solution during training will be proportional to the training time. Algorithm enhancements that shorten training are therefore beneficial. If parts of the digital training circuitry can be turned off throughout the training phase, energy consumption can be reduced even further due to reduced switching activity. \par

In this paper, we introduce a novel variant of the TM learning scheme, incorporating absorbing state TA. That is, once a specific TA state is reached (the absorbing state), we eliminate it permanently. If the TA excluded its literal in the absorbing state, we remove the literal as well. Upon inclusion, on the other hand, we permanently incorporate the literal into its clause. Since the TA is the computational unit of a TM, its removal instantly decreases the training time and energy consumption. Our empirical evaluation on multiple NLP datasets show that the accuracy of our absorbing scheme is equal or better while significantly reducing training time. 

\begin{figure}[htbp]
\centerline{\includegraphics[width=0.4\textwidth]{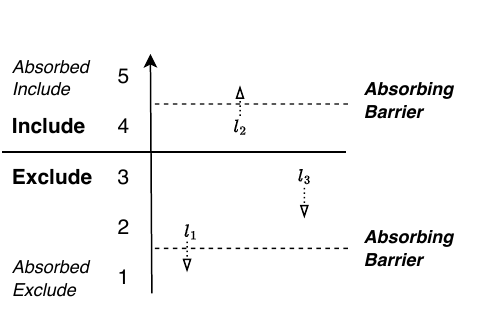}}
\caption{Clause with absorbing states.}
\label{figure:tm_memory}
\end{figure}

\begin{figure}[htbp]
\centerline{\includegraphics[width=0.4\textwidth]{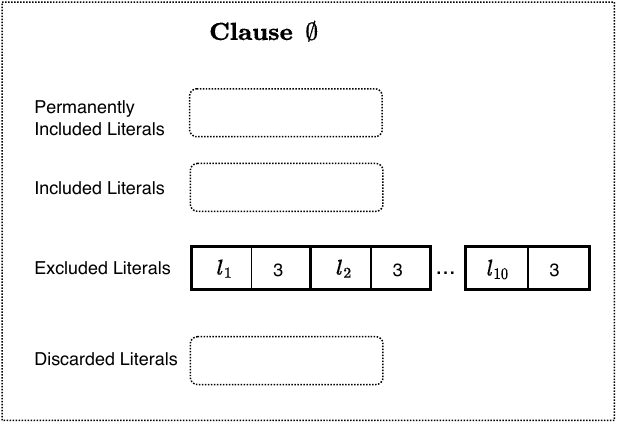}}
\caption{Initialization of sparse clause structure.}
\label{figure:initialization_sparse_tm_structure}
\end{figure}

\begin{figure*}[htbp]
\centerline{\includegraphics[width=0.9\textwidth]{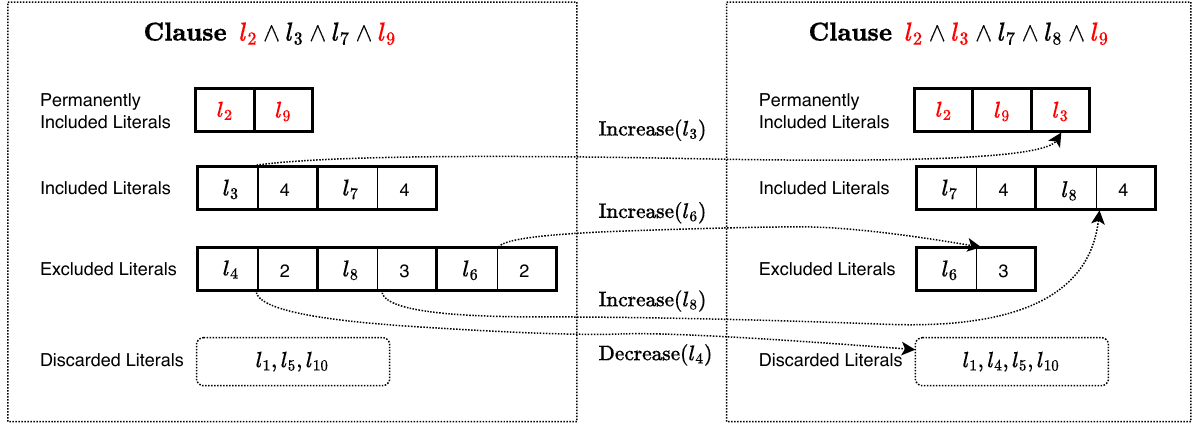}}
\caption{Contracting sparse clause structure.}
\label{figure:sparse_tm_structure}
\end{figure*}

\section{Tsetlin Machine Basics}
A TM consists of a team of two-action TA, each with $2N$ states. For example, an 8-bit TA would produce $256$ states, where states $1$ to $N$ select an ``Exclude'' action and states $N+1$ to $2N$ selects an ``Include'' action (Fig. \ref{figure:ta_vs_absorbTA}, top). Iterative feedback in the form of rewards and penalties drives TA learning. Rewards reinforce the actions with state transitions towards the far end, while penalties suppress the action with state transitions towards the center. In brief, the feedback controls the actions that solve the pattern recognition problem specified below by maximizing the classification accuracy.\par
A TM takes a Boolean input vector $\mathbf{X} =[x_1,\ldots,x_K]$, to be classified into one of the $M$ classes, $Y \in \{y_1, y_2, \ldots, y_{M}\}$. The features are converted into a literal set that consists of the features themselves as well as their negated counterparts: $L = \{x_1,\ldots,x_K,\neg{x}_1,\ldots,\neg{x}_K\}$. To express $\mathcal{S}$ sub-patterns per class, a TM employs $N = M \times \mathcal{S}$ conjunctive clauses, where $\mathcal{S}$ is a user-set parameter. Traditionally, half of the clauses get positive polarity ($+$). The other half gets negative polarity ($-$). For any class $m$, the clauses are denoted $C_n^p$, where $1 \leq n \leq N/2$ is the clause index and $p \in \{-, +\}$ is its polarity. Being conjunctive, the clause becomes:
\begin{equation}
C_n^p(\mathbf{X})=\bigwedge_{l_k \in L_n^p} l_k.
\end{equation}
\noindent Here, $L_n^p$ is a subset of the literals $L$, $L_j^p \subseteq L$. For example, the clause $C_1^+(\mathbf{X}) = \neg x_1 \land x_2$ has index $1$, polarity $+$, and consists of the literals $L_1^+ = \{\neg x_1, x_2\}$. Accordingly, the clause outputs~$1$ if $x_1 =0$ and $x_2 = 1$, and $0$ otherwise.\par

After being evaluated on the input, the clauses' outputs are merged by summation and classified by thresholding using the unit step function $u(v) = 1 ~\mathbf{if}~ v \ge 0 ~\mathbf{else}~ 0$:
\begin{equation}
\textstyle
\hat{y} = u\left(\sum_{n=1}^{N/2} C_n^+(\mathbf{X}) - \sum_{n=1}^{N/2} C_n^-(\mathbf{X})\right).
\end{equation}
As such, the summation of clause outputs produces a majority vote for a classification decision, with positive clauses voting for the class and negative against the class.\par
TM learning involves building conjunctive clauses by means of TA teams. In brief, each literal $l_k$ of each clause $C^p_n$ gets its own TA, which decides whether to Include the literal in the clause. The TA decisions are based on feedback passed in the form of Reward, Inaction, and Penalty. The learning of $TM$ utilizes two types of Feedback: Type I feedback and Type II feedback. Type I Feedback stimulates the formation of frequent patterns, which suppresses false negative classifications. Type II Feedback, on the other hand, increases the discrimination power of the patterns, counteracting false positive classifications. Both types of feedback allow clauses to learn numerous sub-patterns from data. The details of the learning process can be found in~\cite{granmo2018tsetlin}.

\section{Sparse Tsetlin Machine with\\ Contracting Clauses}

Efficient implementation of the Absorbing Tsetlin Machine in software requires adjustments to the internal memory structure. The reason is that we need to be able to  discard arbitrary literals upon reaching the absorbing barrier. Then a literals will no longer be active, as depicted in Fig. \ref{figure:tm_memory}. To facilitate a memory structure that provides the required flexibility for dynamically removing literals and clauses from TMs, we propose to go from a dense to a sparse representation of clauses. We achieve this by storing the active literals and their TAs in lists. Furthermore, we organize the clauses of each class in a hash structure to be able to flexibly add classes as they appear in the data.

 When the clauses are small, which is the typical case for TMs, such a sparse representation has three primary advantages. Firstly, it significantly reduces computation load. Only the active literals are evaluated during inference and inactive TAs are skipped during learning. Secondly, the memory footprint is smaller in comparison to the dense clause and literal memory structures commonly employed. This is because inactive TAs and literals are not stored. Finally, the structure is flexible because one can add and remove classes, clauses, literals, and TAs from the hash table and lists.  

\subsection{Absorbing Tsetlin Automaton}
\begin{figure}[h!!]
\centerline{\includegraphics[width=0.5\textwidth]{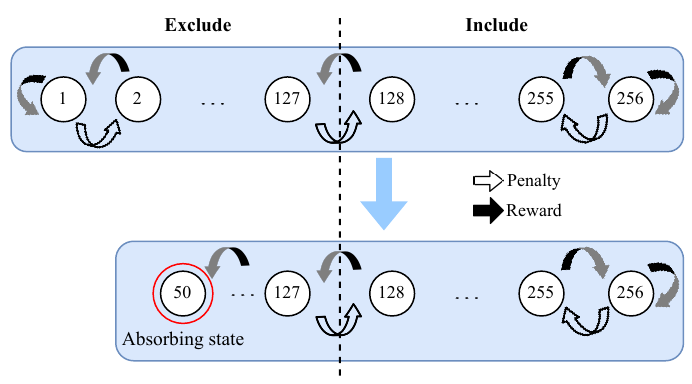}}
\caption{ Standard Tsetlin Automaton \textit{vs} Absorbing Tsetlin Automaton. The absorbing boundary set at the $50^{th}$ state.}
\label{figure:ta_vs_absorbTA}
\end{figure}

%New figure: Standard Tsetlin Automaton vs Absorbing Tsetlin Automaton
While both the Include and Exclude side of the TA can be absorbing, for the sake of simplicity, we here focus on absorbing Exclude. We depict both the standard and the absorbing TA in Fig.~\ref{figure:ta_vs_absorbTA}. In the standard TA depicted in the upper figure, there are $256$ TA states in total, with $1$ to $127$ performing the Exclude action and $128$ to $256$ the Include action.  In the absorbing TA (lower figure), on the other hand, we set the $50^{th}$ state to be absorbing (highlighted in red). As such, the depth of the absorbing state becomes a user-configurable parameter. Whenever an excluded literal reaches the absorbing state, it can never become included in the clause again. It is permanently stuck in the Exclude state.

\subsection{Clauses with Absorbing Tsetlin Automata}

We illustrate the behavior of clauses with absorbing TA in Fig.~\ref{figure:tm_memory}. During the learning process, excluded literals are eliminated from the clauses once they enter the absorbing state, passing the absorbing barrier. As depicted in the figure, we see that literal $l_1$ and $l_3$ are on the Exclude side of the state space. With the absorbing barrier set between state $1$ and $2$, literal $l_1$ is permanently removed as soon as it transitions from state $2$ to state $1$.

\subsection{Absorbing Tsetlin Machine Architecture}

Fig. \ref{figure:initialization_sparse_tm_structure} shows the sparse structure of the absorbing TM clause. Apart from this sparse structure and the accompanying absorbing TA, the architecture is a standard TM architecture. The sparse structure is built on lists that replace the typical dense arrays for storing the TA states. The list structure allows TAs to be removed and added from the structure while learning, introducing a growing and contracting structure. Three lists build up the clause, as follows.
\begin{itemize}
    \item \textbf{Excluded Literals.} This list contains all the TAs that have decided to exclude their literal. The list contains tuples $(l_i, \phi_i)$ consisting of the literal $l_i$ and the state $\phi_i$ of its associated TA, e.g., $3$ in the figure. Initially, all the TAs start in this list, in the state closest to the center on the Exclude side of the clause memory (Fig.~\ref{figure:tm_memory}).
    \item \textbf{Included Literals.} This list has the same structure as the Excluded Literals list, however, it contains all the TAs that have decided to include their literal. The list starts out empty, and will gradually be extended during learning as the TAs decide which literals to include. 
    \item \textbf{Permanently Included Literals.} This list contains the literals whose TAs have been absorbed on the Include side. These literals become a permanent part of the clause.
    \item \textbf{Discarded Literals.} All the literals absorbed on the Exclude side, however, are simply discarded, indicated by the dotted structure in the figure. These literals will never become part of the clause, and we do not need to keep track of these in any manner. 
\end{itemize}

There are several benefits to a list-based structure.
\begin{enumerate}
    \item Computation complexity shrinks as the TAs and their literals are removed from the lists.
    \item Each clause can have its own unique pool of literals to learn from, enabling a more diverse distribution of tasks.
    \item A clause can be quickly evaluated by only traversing the Included Literals lists, which tend to have few elements, in particular with a literal budget~\cite{abeyrathna2023budget}.  
\end{enumerate}   

\subsection{Learning with Absorption}

Fig. \ref{figure:sparse_tm_structure} captures how the sparse list structure supports effective state updates during learning. There are only two types of updates on a TA state: \emph{Increase} or \emph{Decrease} the state.

\textbf{Increase.} The following pseudo-code captures the updating procedure for the sparse structure for Increase:
\begin{enumerate}
    \item If the TA is in the Included Literals list, increase its state by one. If it is now in its absorption state, remove the TA and its literal from the Included Literals list. Then place only the literal at the end of the list of Permanently Included Literals. The update for literal $l_3$ in Fig. \ref{figure:sparse_tm_structure} demonstrates this step (permanently included literals are marked in red). 
    \item If the TA is in the Excluded Literals list, increase its state by one. If the TA's action now has switched to Include, remove the TA and its literal from the list. The removal is carried out by overwriting the tuple with the last element of the list, thereby reducing the size of the Excluded Literals list.  The TA and its literal are finally added at the end of the list of Included Literals. The update for literal $l_8$ in Fig. \ref{figure:sparse_tm_structure} showcases this step. 
\end{enumerate}

\textbf{Decrease.} The Decrease operation also discards TAs and their literals:
\begin{enumerate}
    \item If the TA is in the Exclude Literals list, decrease its state by one. If it now is in its absorbing state, remove the TA and its literal from the list of Excluded Literals, thereby discarding them. The discarding of literal $l_4$ in Fig. \ref{figure:sparse_tm_structure} exemplifies this step. 
    \item If the TA is in the Included Literals list, decrease its state by one. If the TA's action now has switched to Exclude, remove the TA and its literal from the list. The removal is carried out by overwriting the tuple with the last element of the list, thereby reducing the size of the Include Literals list. The TA and its literal are finally added at the end of the Excluded Literals list.
\end{enumerate}

%%%%%%%%%%%%%%%%%%%%%%%%%
\section{Hardware Solutions for Sparse Contracting Tsetlin Machine}

The dynamic power, $P$, consumed by a digital circuit with load capacitance $C$, operating frequency $f$, supply voltage $V_{S}$, and with an activity factor $\alpha$ (transitions per clock cycle) is given by $P=0.5 \times C \times V_{S}^2 \times f \times \alpha$~\cite{Dally}. Turning off TAs during training, due to reached absorbing states, will reduce the switching activity of these modules, and also the clause logic controlled by these TAs. This will lead to reduced power consumption. The exact savings will depend on the TM configuration and the dataset. 

\subsection{Reduction of Switching Activity}

In a dedicated hardware solution, such as a Field Programmable Gate Array (FPGA) or an Application Specific Integrated Circuit (ASIC), a TA can typically be implemented as a binary up-down counter. %The Most Significant Bit (MSB) of the counter can be applied as the Include/Exclude signal.

It is relatively straightforward to implement absorbing states in such a TA. In more detail, it will require an extra logical check in the hardware description (e.g., in VHDL or Verilog) of the TA, which disables further updating if the absorbing state has been reached. 
%In its simplest form such disabling of a TA can be performed by controlling the enable signal to the \mbox{D flip-flops} that  constitute the TA. 
% Fig. \ref{figure:ta_vs_absorbTA} shows the principle for this. Here the Include states are from 128 and above, while the lower states imply Exclude. The upper part of the figure shows a standard TM, while the lower part shows a TM with state 50 as absorbing. In Verilog/VHDL one can easily define the minimum and maximum state numbers and whether they are absorbing or not. 

\begin{figure}[h!!]
\centerline{\includegraphics[width=0.4\textwidth]{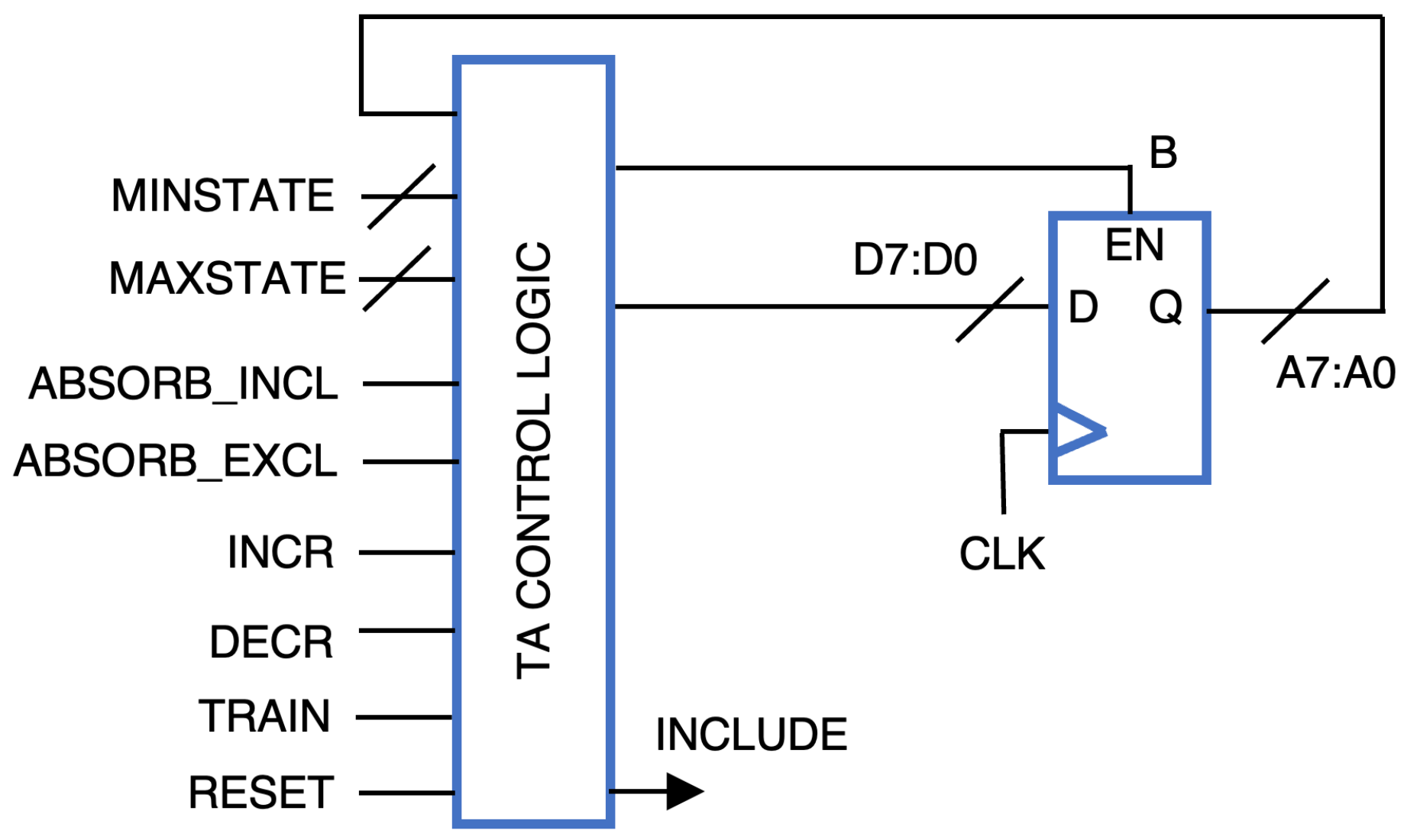}}
\caption{Block diagram example of a single TA state machine. The next state is defined by the TA control logic.}
\label{figure:ta_state_machine_block_diagram}
\end{figure}

Fig. \ref{figure:ta_state_machine_block_diagram} shows a possible circuit implementation of an eight-bit TA. Here the states from 0-127 imply \textit{Exclude} and 128-255 imply \textit{Include}. The TA control logic, together with the various TA input signals and the current TA state (A7:A0), determine the next state (D7:D0) of the TA, which will be available at the output of the D flip-flops (A7:A0) at the next clock (CLK) low-to-high transition. The TA output signal INCLUDE is typically derived from the MSB (A7), either directly (active low) or inverted. If RESET is high, the next state of the TA is typically set to 127. During training, other circuit modules will decide if the state of the TA should be increased (INCR is set high) or decreased (DECR is set high), or if the state should remain unchanged (INCR and DECR are both low). 

The inputs MINSTATE and MAXSTATE here define the minimum and maximum TA states respectively. \mbox{ABSORB\_INCL} and \mbox{ABSORB\_EXCL} define if the maximum and minimum states are absorbing or not. If absorbing, the control logic will ensure that no changes in D7:D0 will be made if an absorbing state has been reached. In addition, this will set the enable signal B low. (B will also be set low if the TRAIN signal is low, indicating that training is not ongoing.) Furthermore, if needed by the architecture, status signals can be generated by the TA control logic that indicates if an absorbing Exclude or Include state has been reached.

\begin{table*}
\centering
\caption{Performance comparison of Tsetlin Machine with different absorbing states. The results reported are in \% (seconds) from last $25$ epochs.}
    \begin{tabular}{c|c|c|c|c|c|c}
    \hline
    \multirow{2}{*}{Datasets} &  \multicolumn{6}{c}{ Absorbing State }\\ \cline{2-7}
    & $0$ & $25$ & $50$ & $75$ & $100$ & $125$\\
    \hline
    TREC & 91.22 (50.99) & 91.57 (8.03) & \textbf{91.61} (5.77) & 91.24 (4.04) & 90.0 (3.23)& 74.87 (2.86)\\
    MPQA & 67.42 (186.07) & 68.10 (5.87) & 70.61 (5.54) & \textbf{73.85} (5.09) & 68.59 (5.16) & 58.09 (5.07)\\
    SUBJ & 79.24 (308.16) & 81.04 (6.70) & \textbf {81.78} (6.06) & 81.29 (5.75) & 79.84 (5.53) & 50.60 (2.76)\\
    CR & \textbf {75.95} (48.36) & 75.91 (2.63) & 75.15 (2.11) & 72.61(1.90) & 71.72 (1.80) & 66.08 (1.62)\\
    R8 & 95.37 (20.86)& 95.62 (12.15) & 95.72 (9.07) & \textbf {95.84} (6.39) & 95.24 (3.97) & 84.77 (3.31)\\
    R52 & 88.98 (11.34) & 89.07 (10.09) & \textbf {89.53} (10.23) & 89.19 (9.61) & 89.47 (8.25) & 79.34 (7.01)\\
    \hline
    \end{tabular}
\label{results_epochs}
%\squeezeup
\end{table*}

In many cases, a significant part of the digital switching power stems from the clock tree. Therefore, one will achieve the greatest reduction in switching activity if clock-gating is used, and even more, if power-gating is implemented for turning off TAs. However, controlling this with very fine granularity might result in sub-optimal solutions, because extra logic and routing will be needed. Possibly, for larger groups of TAs, if the whole group has reached absorbing states, such logical control of the clock or power gating could be performed. 

\subsection{Reduction of Training Time}

Turning off TAs from further updating, can lead to reduced training time also for dedicated hardware solutions. However, this is dependent on the circuit architecture and the exact implementation of TAs, and the updating method. 

For a sequential TA updating approach, e.g., implemented with a program running on a low-power microcontroller, being able to skip updating of TAs in absorbed states can lead to a significant training time reduction. 

On the other hand, a dedicated TM hardware accelerator, with a highly parallel architecture, training time reduction might not be significantly affected by skipping such updating of TAs. For instance, a solution where all TAs belonging to a clause are updated in parallel, during a single clock cycle, will not achieve training time reduction. Nevertheless, the reduced switching activity from disabled single TAs will reduce power consumption. 

Obviously, the technique of disabling TAs during training after they have reached an absorbing state will only affect the training mode. For inference, the TA decisions (Include or Exclude) from the training phase are kept, typically in a register constituted by the MSBs of the TAs. To reduce power consumption, the parts of the TAs that are not needed during inference, as well as other training-specific modules, can be disabled.

%%%%%%%%%%%%%%%%%%%%%%%%%
\section{Empirical Results}
In this section, we provide the implementation details and evaluate the performance of the TM with absorbing TAs. In particular, we demonstrate how absorbing at different states influences the training time while maintaining accuracy. 

\subsection{Datasets}
We run experiments on six publicly available text classification datasets.
\begin{itemize}
    \item TREC-6~\cite{chang2002system} is an open-domain, fact-based question classification dataset.
    \item MPQA~\cite{wilson2005recognizing} is a dataset for detecting opinion polarity.
    \item CR~\cite{ding2008holistic} is a customer review dataset with each sample labeled as positive or negative.
    \item SUBJ~\cite{pang2004sentimental} concerns classifying reviews as subjective or objective.
    \item R8~\cite{debole2005analysis} is a subset of the Reuters-21578 with $8$ classes.
    \item R52~\cite{debole2005analysis} is a subset of the Reuters-21578 with $52$ classes.
\end{itemize}

\begin{table}
\centering
\caption{Performance comparison of Tsetlin Machine with different absorbing states and subsampling rate in R8 dataset. The results reported are in \% (seconds) from last $25$ epochs.}
    \begin{tabular}{c|c|c|c|c}
    \hline
    \multirow{2}{*}{States} &  \multicolumn{4}{c}{ Subsampling }\\ \cline{2-5}
    & $0.1$ & $0.3$ & $0.6$ & $0.9$ \\
    \hline
    0 & 95.66 (36.56)& 95.46 (89.55)& 95.58 (162.03) & \textbf {95.99} (243.12)\\
    25 & \textbf{95.73} (25.26) & 95.55 (62.89) & 95.52 (119.38) & 95.61 (172.22)\\
    50 & 95.49 (23.96) & \textbf{95.77} (57.56) & 95.68 (106.42) & 95.70 (152.94)\\
    75 & 95.47 (17.88) & 95.44 (41.55) & \textbf{95.63} (76.97) & 95.54 (111.218)\\
    100 & 95.30 (11.26) & 95.46 (22.46) & \textbf{95.79} (38.91) & 95.66 (55.13)\\
    125 & 80.19 (4.41) & 82.61 (5.04) & 85.05 (5.95) & \textbf{88.51} (6.71)\\
    \hline
    \end{tabular}
\label{results_subsample}
%\squeezeup
\end{table}

We utilized the predefined train and test splits for all the datasets.

\subsection{Implementation Details}

 The TM model is initialized with the following parameters: number of clauses~($N$), voting margin~($T$), specificity~($s$), maximum included literals, literal sampling, and an absorbing state. The performance of TM is initially optimized for each dataset using $N$, $T$, and $s$. We then vary the absorbing state, while also sampling literals to equip each clause with a random subset of the literals. In the first experiment, we change the absorbing state from $0$ to $125$ and record the accuracy, training time, and absorption rate. In the second experiment, we measure performance when varying the absorbing state for literal sampling values ranging from $0.1$ to $0.9$. All experiments were conducted on a NVIDIA DGX-2 server with dual Intel Xeon Platinum 8168, 2.7 GHz, 16$\times$ NVIDIA Tesla V100 (32~GB), and Ubuntu 18.04 LTS x64.

\begin{figure}
\centerline{\includegraphics[width=0.5\textwidth]{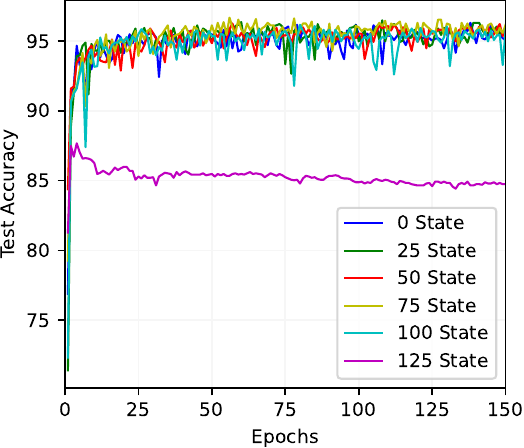}}
\caption{Test accuracy \textit{vs} absorbing states per epoch in R8 dataset.}
\label{figure:acc_vs_absorb}
\end{figure}

\subsection{Results and Discussions}

In this section, we present and discuss our findings regarding TMs with absorbing TA. We start with state $25$ as absorbing Exclude and progressively shift it up to state $125$. We then record the resulting accuracy and training time for each of the six datasets. The TM performance for the different absorption states is shown in Table~\ref{results_epochs}. We first observe that only dataset $CR$ is marginally better without absorption. For three out of six datasets, we record improved accuracy with absorbing state $50$, while two datasets showed improved accuracy with absorbing state $70$. Additionally, we measure a significant reduction in training time for all of the datasets with increasing absorption state. For instance, the SUBJ and CR datasets showed training time reduction of $51 \times$ and $23 \times$, respectively, while achieving equal or increased accuracy for absorbing state $50$. Notably, we found that the absorbing state at $125$ led to the worst performance, as all the literals are absorbed only two states away from the center. Overall, our results suggest that a trade-off between accuracy and training time can be achieved by selecting an absorbing state between $50$ and $75$.\par

\begin{figure}
\centerline{\includegraphics[width=0.5\textwidth]{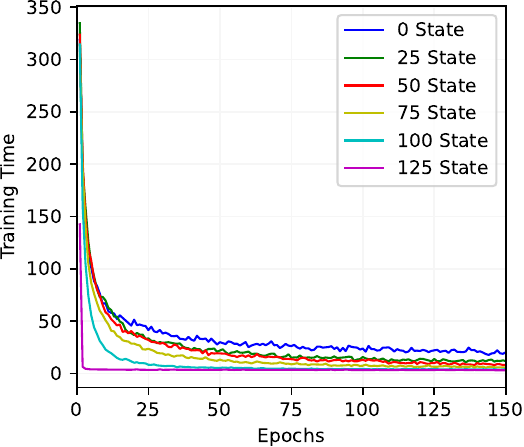}}
\caption{Training time \textit{vs} absorbing states per epoch in R8 dataset.}
\label{figure:time_vs_absorb}
\end{figure}

\begin{figure}
\centerline{\includegraphics[width=0.5\textwidth]{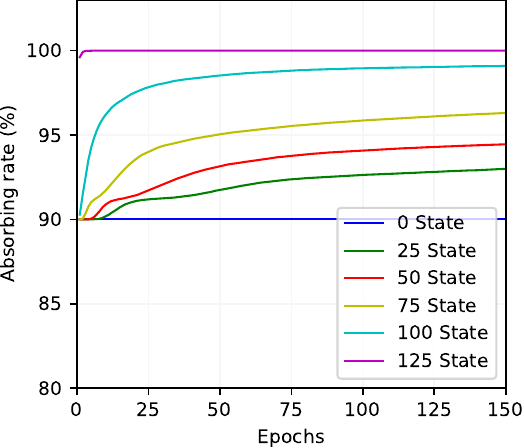}}
\caption{Absorbing rate \textit{vs} absorbing states per epoch in R8 dataset.}
\label{figure:acc_vs_rate}
\end{figure}

We further investigate how test accuracy and training time evolve over $150$ epochs on the R8 dataset. The results are plotted in Fig.~\ref{figure:acc_vs_absorb} and Fig.~\ref{figure:time_vs_absorb}. We notice that we are able to get competitive accuracy with absorbing states from $0$ to $100$, typically being attained after around $25$ epochs. And for absorbing state $125$, the accuracy does not improve. We also noted a decrease in training time with increasing absorbing state, particularly in the first $25$ epochs, where the decrease was exponential. To better understand the relationship between the absorbing state and the rate of literals being absorbed, we recorded the absorbing rate of literals across all absorbing states for $150$ epochs. We hypothesized that when the absorbing states are near the center, i.e., absorbing states $125$, $100$, and so on, the literals are most frequently absorbed. This is supported by our findings in Fig.~\ref{figure:acc_vs_rate}. \par

We now present a second experiment conducted on the R8 dataset to investigate the impact of initializing the clauses with only a subset of the literals available. Specifically, we gradually increased the fraction of literals available from $0.1$ to $0.9$ in intervals of $0.3$. We record the accuracy and the training time for each absorbing state and the findings are reported in Table~\ref{results_subsample}. Our results show that increasing the pool of literals available to each clause generally leads to an improvement in accuracy, with the effect being more significant with absorbing states $0$ and $125$.  However, We also observe that the training time significantly increases with an increase in literal samples, while it decreases with absorbing states. This means that absorbing state $0$ with a literal sample of $0.9$ is the most computationally inefficient (i.e., training time is $6.75 \times$ more than literal sample $0.1$). Moreover, as the absorbing states increase, the training time for $0.9$ samples decreases, reaching a minimum of only $1.52 \times$ that of a literal sample $0.1$ for $125$ absorbing state. This observation can be explained by the fact that as the absorbing state increases, the literal tends to get more absorbed, thus reducing the overall training time even with the introduction of more literal samples at the beginning of the training.

\section{Conclusion}
In this paper, we propose a novel contracting TM with clauses that utilize absorbing TAs. Our hypothesis is that by removing excluded literals from the learning process, we can significantly reduce computation time during training. This hypothesis is supported by extensive experiments on six NLP datasets. Our experimental results demonstrate that the proposed TM achieves the same or even better accuracy with significantly reduced training time, especially when different absorbing TA states are set. Additionally, we conclude that the impact of adding more literals at the beginning decreases as the number of absorbing states increases.

\section*{Acknowledgment}
This work is supported in part by the project ``Logic-based Artificial Intelligence Everywhere: Tsetlin Machines in Hardware'' and funded under grant number 312434 of the Research Council of Norway.
%\section*{References}
\bibliographystyle{IEEEtran}
\bibliography{biblio}
\end{document}